\documentclass[structabstract]{aa}  
\usepackage{graphicx}
\usepackage{txfonts}
\usepackage[sort&compress]{natbib} 
\begin{document}
     \title{An 8-mm diameter Fiber Robot Positioner for Massive Spectroscopy Surveys}

   \author{N. Fahim
          \inst{1},
            F. Prada
          \inst{2,3,4},
          J. P. Kneib 
          \inst{5,6},  
           J. S\'anchez
          \inst{4},
          P. H\"orler 
          \inst{7},
          M. Azzaro 
          \inst{4},
          S. Becerril
          \inst{4},
           H. Bleuler
          \inst{7},
           M. Bouri
          \inst{7},
          J. Castano
          \inst{8},
          J. Garrido
          \inst{1},
          D. Gillet
          \inst{7},
          G. Glez-de-Rivera
          \inst{1},
          C. G\'omez
          \inst{2},
          M.A. G\'omez
          \inst{8},
          A. Gonzalez-Arroyo
          \inst{2,10},
          L. Jenni
          \inst{7},
          L. Makarem
          \inst{7},
          G. Yepes
          \inst{10}
          X. Arrillaga 
          \inst{11},
           MA. Carrera
          \inst{11},  
          R. Diego
          \inst{11},
          M. Charif
          \inst{12},          
          M. Hug
          \inst{13},          
          C. Lachat
          \inst{13},          
          }

   \institute{
     Grupo de Investigaci\'on HCTLab, Escuela Polit\'ecnica Superior, Universidad Aut\'onoma de Madrid, Spain
	\and 
   Instituto de F\'isica Te\'orica, (UAM/CSIC), Universidad Aut\'onoma de Madrid,  Cantoblanco, E-28049 Madrid, Spain
		\email{f.prada@csic.es}
	\and
   Campus of International Excellence UAM+CSIC, Cantoblanco, E-28049 Madrid, Spain 
	\and
   Instituto de Astrof\'isica de Andaluc\'ia (CSIC), Glorieta de la Astronom\'ia, E-18080 Granada, Spain 
	\and
    Laboratoire d'Astrophysique, Ecole Polytechnique F\'ed\'erale de Lausanne, Observatoire de Sauverny, CH-1290 Versoix, Switzerland
    \email{jean-paul.kneib@epfl.ch}
    \and
	Aix Marseille Universit\'e, CNRS, Laboratoire d'Astrophysique de Marseille, UMR 7326, 13388, Marseille, France 
	 \and
    Laboratory of Robotic Systems (LSRO), Ecole Polytechnique F\'ed\'erale de Lausanne (EPFL), 1015 Lausanne, Switzerland 
    \and
    Escuela T\'ecnica Superior de Ingenieros Aeron\'auticos (ETSIA), Universidad Polit\'ecnica de Madrid, Spain
    \and
    Grupo de Electr\'onica y Semiconductores, UAM, Madrid, Spain 
    \and
    Departamento de F\'isica Te\'orica, Universidad Aut\'onoma de Madrid, Cantoblanco, 28049, Madrid, Spain
    \and 
    AVS (Added Value Solutions), Elgoibar, Gipuzkoa, Spain
    \and
    FAULHABER, miniature drive systems, Switzerland 
	\and
	MPS Micro Precision Systems AG, PO Box 8361, CH-2500 Biel/Bienne, Switzerland
             }

   \date{Received -, 2014; accepted -, 2014}

 
  \abstract
   {
   Massive spectroscopic survey are becoming trendy in astrophysics and cosmology, as they can address new fundamental knowledge such as
   Galactic Archaeology and probe the nature of the mysterious Dark Energy. To enable massive spectroscopic surveys, new technology
   are being developed to place thousands of optical fibers at a given position on a focal plane. These technology needs to be: 
   1) accurate, with micrometer positional accuracy; 2) fast to minimize overhead; 3) robust to minimize failure; and 4) low cost.
   In this paper we present the development of a new 8-mm in diameter fiber positionner robot using two 4mm DC-brushless  gearmotors, 
    developed in the context of the Dark Energy Spectroscopic Instrument.
    This development was conducted  by a Spanish-Swiss (ES-CH) team led by the Instituto de F\'isica Te\'orica (UAM-CSIC) 
    and the Laboratoire d'Astrophysique (EPFL), 
    in collaboration with the AVS company in Spain and the Faulhaber group (MPS \& FAULHABER-MINIMOTOR) in Switzerland.
		}
   {The meachanical concept,  DC-brushless motor properties, and the final performance of a prototyped unit is presented.}
   {Performance and verification tests were conducted with a fiber view camera-based optical set-up and using an automatic algorithm.}
   {The prototype build is mechanically robust and reliable, and its control electronics ensure a very firm system with an xy
   positional accuracy better than $5\mu$m.}
   {In this paper, we validate the concept of our advanced 8-mm fiber robot positioner prototype, as well as demonstrate that
   it can meet the requirements of the DESI project. Such efficient gearmotor fiber positionner robotic system can be adapted 
   to any future massive fiber-fed spectrograph instrument.}
	
   \keywords{astronomical instrumentation -- mechatronics -- fiber robot positioner -- fiber-fed spectrographs}
   \authorrunning{-N.Fahim, et al.}
   \maketitle
%
%
%
%
%
%
\section{Introduction}

The recent Sloan Digital Sky Survey (SDSS) measurements of the 
Baryonic Acoustic Oscillation (BAO) peak in the distribution of galaxies
\cite{Eisenstein2005}, \cite{Anderson2012} and in the Ly-$\alpha$ forest of
distant quasars \cite{Delubac2014} are drawing new and important constraints 
on the cosmological world model. These successful measurements are 
paving the way for even more massive spectroscopic surveys that will
improve significantly the number of targeted object in order to 
derive more stringent constraints on the cosmological world model with a particular
focus on the characterisation of the nature of Dark Energy - or at least its equation of state.

However to enable such massive survey new technology is needed to increase
significantly the multiplexing capabilities of multi-object spectrograph on 4-8m class telescopes. 
There are two main types of multi-object instruments: multi-slit spectrographs and fiber-fed spectrograph. Both offer advantages and disadvantages, but the most versatile type for object collection is the fiber-fed approach. 
For all the fiber-fed spectrographs, the key issue to solve  is the accurate 
positioning of the many optical fibers at the location of the targeted objects 
(star, galaxy or quasar) at a given sky position.
For each sky position, the placement of the fibers is different. Therefore, 
the fiber positioning system has to be capable to adjust the position of fibers
to the new targeted positions within a short amount of time
(ideally faster than the detector readout time - typically less than one minute; 
or if done in parallel of the actual observation faster than the integration time).
Different strategies have been used in previous instruments (e.g. see \cite{Smith2004} for a general review, or \cite{Haynes2006} for a review on positioner technology):
1) for the SDSS spectroscopy surveys, fibers are placed by hand on a set of cartridges that are prepared during day time; 
2) for the AAT/AOmega spectrograph, each fiber are sequentially picked and placed 
at their targeted position on a plate by a robotic system, while in parallel
another plate is being observed ensuring a good duty cycle;
3) for LAMOST (\cite{Zhang2008}), Subaru/FMOS  (\cite{Moore2002}) an array of positioners places all the fibers simultaneously at the targeted positions, using back-illumination of the fibers
to accurately place fibers at the position of the targets. Such concept is the most versatile, as targets can be decided just before observations and the
reconfiguration time is extremely short (typically less than one minute).

The new massive spectroscopy projects are almost all adopting the array of positioners as their fiber positioning system, we can list: 
 SIDE (Azzaro et al. 2010), 
  PFS/SuMIRe\footnote{ http://sumire.ipmu.jp/en/2652} with the
  COBRA (\cite{Fisher2009}) positioner, 
 4MOST\footnote{http://www.4most.eu/},
 MOONS\footnote{http://www.roe.ac.uk/~ciras/MOONS/VLT-MOONS.html},
 and DESI\footnote{http://desi.lbl.gov/} (\cite{Silber2012}; and this work). 
These experiments will obtain optical/IR spectra for 
millions of galaxies, quasars and stars. 

In this context, the group lead by F. Prada, in collaboration with the company AVS in Spain, has a long heritage and a valuable experience on developing precision fiber positioning mechanisms. We built a 29mm-pitch positioner for the 10m-GTC telescope in the context of the SIDE instrument proposal (\cite{Prada2008}), which (R\&D) expertise has allowed AVS to build the MEGARA fiber robot positioner (\cite{Gil2014}). We also developed a 12mm-pitch $\theta-\phi$ fiber positioner competing prototype for BigBOSS (\cite{Fahim2013}) that deliver excellent performance results, as presented in several BigBOSS/DESI collaboration meetings. This device, built late 2012 by the company AVS in partnership with IAA-CSIC and EPS-UAM, uses 6mm geared stepper motors. 

To go beyond the above development and reach a smaller size device of $\sim$8mm 
diameter (meeting the latest DES requirements) we have work on a new concept
based on a Spanish-Swiss collaboration led by UAM and EPFL an including
the following companies: AVS in Spain and the FAULHABER group (MPS and FAULHABER-MINIMOTOR) in Switzerland.
The extended Spanish-Swiss group (hereafter ES-CH) has collaborated to design and manufacture a 10.4mm-pitch positioner based on the DESI science and technical requirements. The main objective was to reduce the size of the device, improve its mechanical robustness, but keeping the excellent performances of previously
developed positioners.
The result of this project is presented in this paper, which describes the positioner concept design, and show the performance results of the present ES-CH 10.4-pitch fiber positioner.

We organise the paper as follows. In section 2, we describe the positioner focal plane environment. The mechanical requirements are described in section 3, and the positioner concept design is described in section 4. In section 5 we present the properties and performance of the 4-mm DC-brushless gear-motors used in the positioner. Performance and test results of the positioner are discussed in section 6. We conclude in section 7.

%
%

\section{The context of the new fiber positioner}

\subsection{The DESI project}

The Dark Energy Spectrograph Instrument (DESI) project (\cite{Schlegel2011} and \cite{Schlegel2008}) aim to develop a new massive spectrocopy instrument
to probe the nature of Dark Energy. This project funded largely by the US
Department of Energy will use the NOAO 4-meter Mayall Telescope (Kitt Peak, Arizona, USA) on which will be mounted a fiber-fed spectrograph system. 

\subsection{The Focal Plane Array}

\begin{figure}[!h]
   \centering
   \includegraphics[width=7cm]{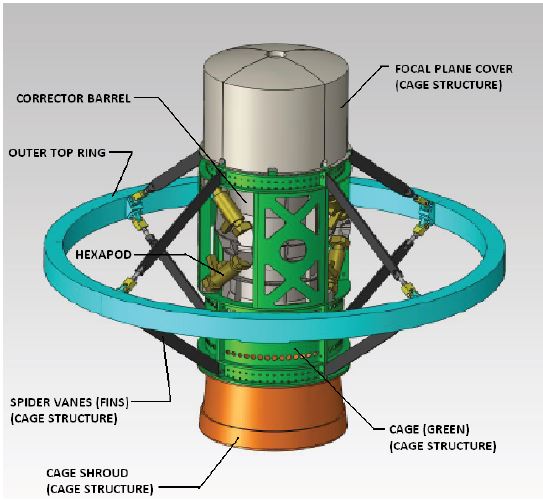}
 \caption {The DESI prime focus optical structure at the NOAO 4-meter Mayall telescope, which contains the focal plane with the positioner array (Image Credit: DESI).}
 \label{focal0}
  \end{figure}
  
\begin{figure}[!h]
   \centering
   \includegraphics[width=7.5 cm]{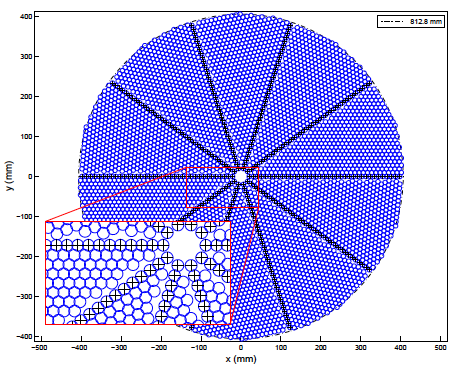}
 \caption {Focal Plate layout. Down-left: front zoom view of the Focal Plate petals mounted on the Integration Ring with some additional support.}
         \label{focal1}
  \end{figure}

Mounted at the new prime focus optical structure (see Figure \ref{focal0}), an array of 5,000 fiber robot positioners (following the $\theta$-$\phi$ philosophy) will be installed  on a curved focal plate (\ref{focal1}). This system will be able to control individually the position of 5,000 optical fibers with an accuracy better than 5 $\mu$m. 
  
Each positioner can place a fiber at any location over a disc which will be called {\it patrol disc}. Adjacent patrol discs overlap so that all the positions on the focal plane can be reached by at least one positioner. Of course, some parts of the focal plane can be reached by more than one positioner, leading to the possibility of collisions between the arms of adjacent robots. To avoid this collisions, moving in parallel and in minimum time, we have developed a motion-planning algorithm for fiber positioners based on a novel decentralised navigation function (\cite{Makarem2014}). The functionality and performance of this algorithm is described in section \ref{algorithms}.
	
\begin{figure}[!h]
  \centering
   \includegraphics[width=7.5cm]{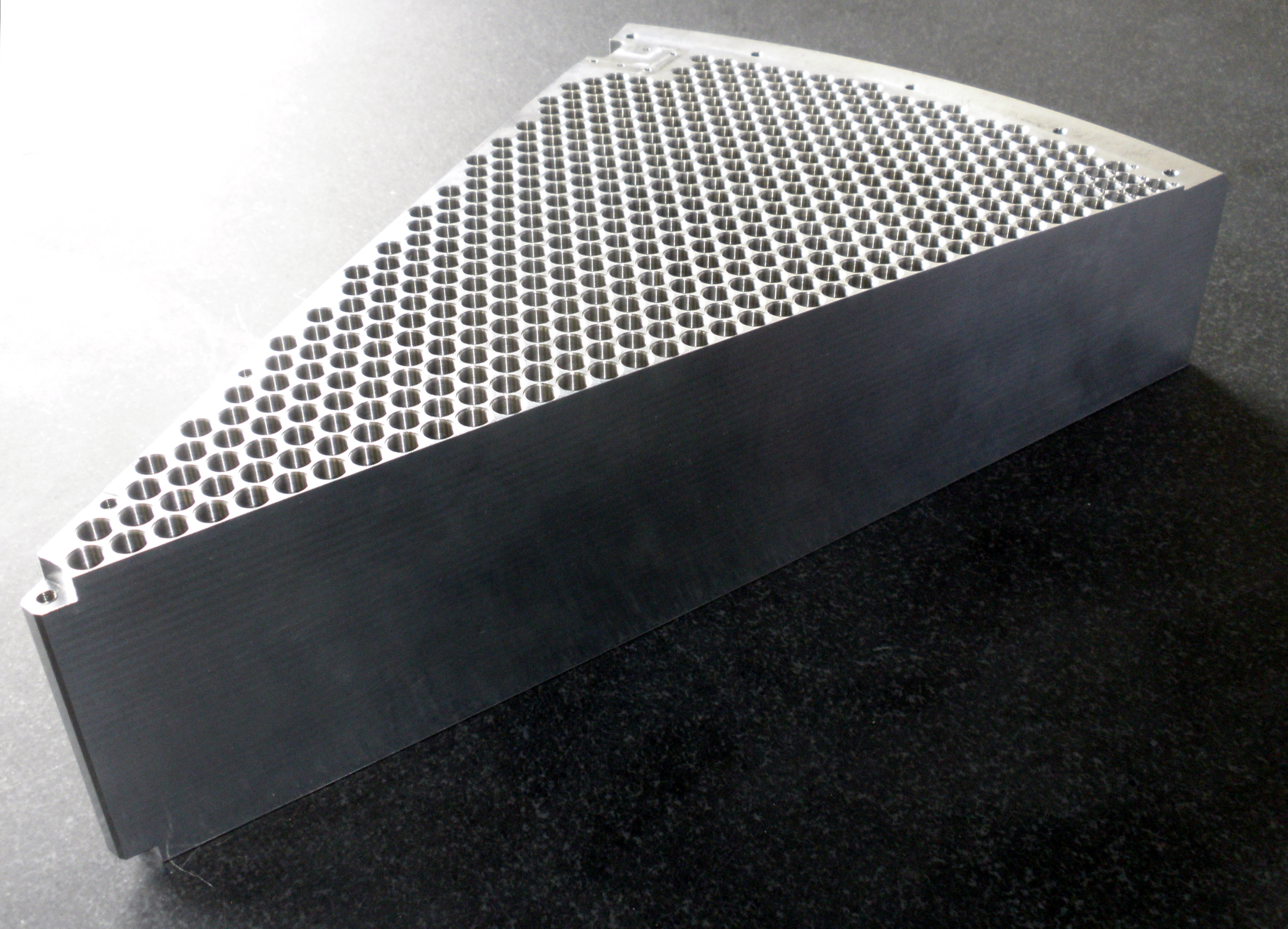}
      \caption{An individual petal, containing 507 positioner holes. 10 of this petals will conform the entire DESI focal plate structure (Image Credit: IAC). 
              }
         \label{petal}
   \end{figure}

\begin{figure}[!h]
  \centering
   \includegraphics[width=7.5cm]{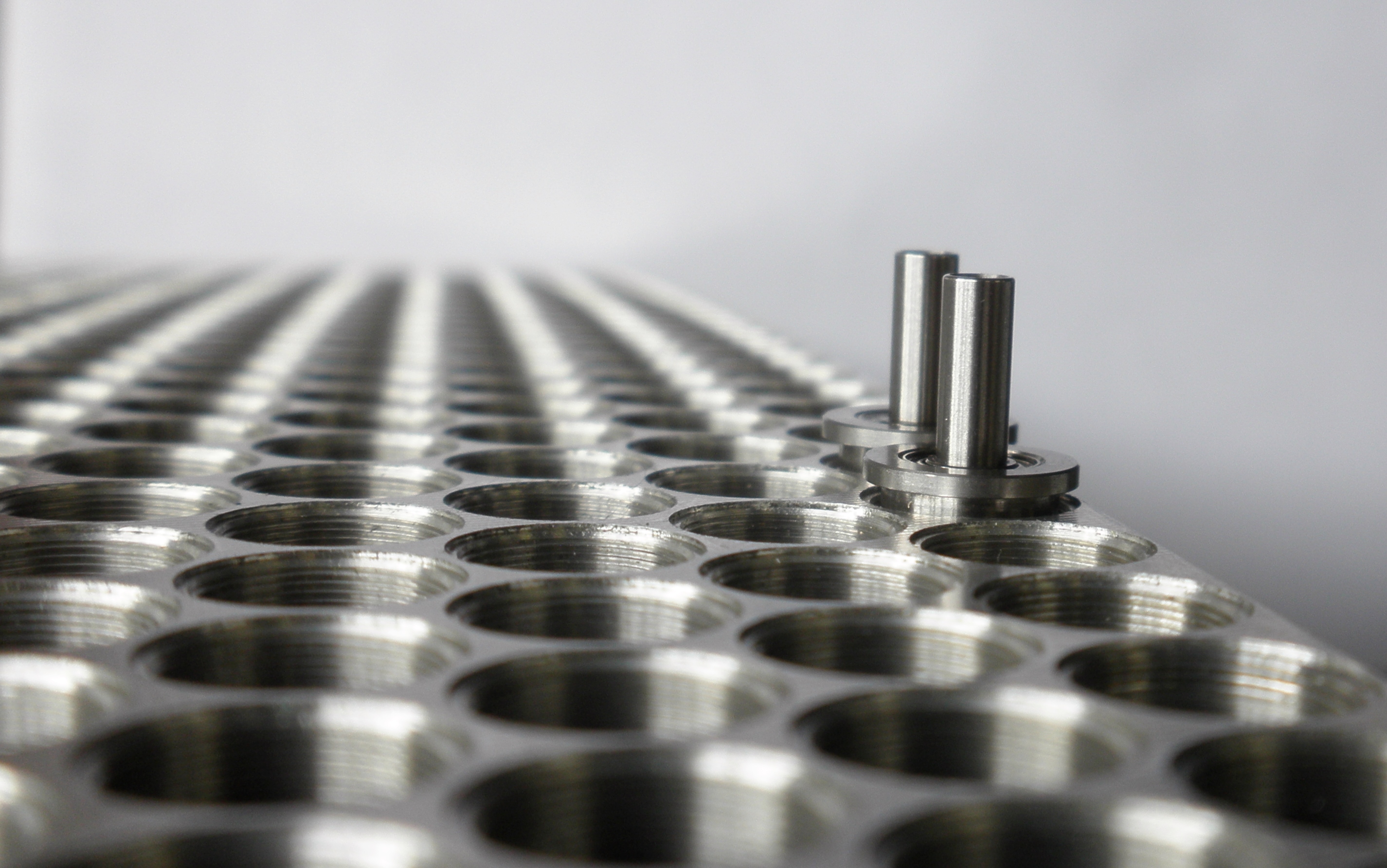}
      \caption{Detail of two dummy positioners installed in two holes (Image Credit: IAC).
              }
         \label{petal_detail}
   \end{figure}

To ensure the micro-metric placing of all the positionners in the focal plane, and to adapt the fiber to the optical curved field, a sectored focal plate has been developed (see Figure \ref{focal1}). This structure is divided into 10 sections called petals, that contain  the positioners holes and a number of fiducial fibers  used to calibrate the metrology of the entire system. Figure \ref{petal} and \ref{petal_detail} shows one petal of this structure as built by the Instituto de Astrof\'isica de Canarias (IAC) in collaboration with LBNL, the Instituto de Astrof\'isica de Andaluc\'ia (IAA-CSIC), and the Instituto de F\'isica Te\'orica UAM/CSIC. Different mechanical interfaces to placing the robots in the focal plate holes have been tested in order to find out the proper way to maximise the accuracy of the whole system.  

This array of fiber robot positioners offers many advantages. First of all, the reconfiguration times between two observing configuration are extremely short. The robots move in parallel all the 5,000 fiber following anti-collision trajectories.
The trajectories are computed off-line following the avoidance collision algorithm.
Therefore, the reconfiguration time of the fiber array is minimised (all the 5,000 fibers are positioned in less than 45 seconds), optimising the sky observation periods. In practise, each positioner is controlled by a dedicated control hardware attached
to the device. This philosophy provides an extremely robust, scalable and easy to maintain system, allowing to control each positioner individually, as well as any hypothetical failures of each robot (motor faults, collision errors, hardware failures, mechanic problems ...). 

Although our team has worked on all the fiber positioner and focal plate aspects, in this paper we will focus on the description of one individual positioner, describing its main characteristics, and reporting the most relevant performance results on positioning accuracy, reconfiguration times, mechanical advantages, as well as providing the results of the characterisation of the adopted 4-mm DC-brushless FAULHABER motors. A future work will be devoted to the focal plate and will be published somewhere else.

%
%
%
%
%

\section{Mechanical requirements}

%
  
We list in Table 1 the most relevant requirements of the DESI fiber robot positioner. 
This paper focuses on the most critical aspects related to the position errors,
which affect directly the position accuracy of the fiber within its patrol disk. 
The X-Y maximal error, the  X-Y RMS error, the Z maximal error and the angular tilt are the main values qualifying the position accuracy.
Nevertheless, other important parameters are also discussed, in particular the power consumption, which can impact the signal-to-noise ratio if important temperature gradient are created, producing turbulent air flows near the focal plate.

 \begin{table}[!h]
         \centering
         \begin{tabular}{l l}
            \hline
            \noalign{\smallskip}
            Item & Value\\
            \noalign{\smallskip}
            \hline
            \noalign{\smallskip}
            Packing Pattern Geometry & Circular\\
            Distance between positioner centers & $10.4$mm \\
			Patrol zone outer radius & $6$mm\\
            RMS XY Error & $< 5\mu$m \\
            Max XY Error & $< 10\mu$m \\
            Z max defocus error & $< 6.4\mu$m \\
            Max angular tilt & $\pm 0.1{}^{\circ}$ \\
            Max reconfiguration time & $20$ sec \\
			Power while active & $1.204$W\\
			Power while sleep & $0.2$mW\\
            Fiber bend radius & $> 50$mm\\
			Operational temperature range & $-10 \,\, +30 \,{}^{\circ}$C \\
            Mass of the positioner  & $50$g \\
            Lifetime  & $10$ years\\
            \noalign{\smallskip}
            \hline
						
         \end{tabular}
				 \caption{Positioner basic requirements.}
         \label{positionerRequirements}
   \end{table}

Z position errors of the fiber tip produces a de-focus error which translates into a large spot not fitting the fiber's core, so a loss of energy is produced.
Angular tilt errors lead to a Focal Ratio Degradation (FRD) at the fiber's exit, with the optical performance being downgraded.

We list below the different elements that can affect the  positioner accuracy:

\begin{itemize}
\item[1.] {\bf Size of the positioner parts:} The size of the parts
are not exactly what is defined in the design (due to errors at the production level, or due to temperature variation). These differences induces an error in the position of the fiber. The most critical parts are the length of the two positioner arms. 
\item[2.]{\bf Assembly imperfections:} The dimensions of the assembled positioner can also differ from what is defined in the design due to assembly imperfections (misalignment of parts for gluing, or unevenly tightened screws). The most critical assembly criterion is the alignment of the two rotation axes. 
\item[3.]{\bf Motion transmission irregularities:} mechanical
imperfections of the gears involved in the motion transmission can create irregularities in some areas of the positioner patrol disc.
\item[4.] {\bf Focal plate accuracy errors:} Such errors may be introduced by the focal plate hole-placing design or production. An angular deviation of the housing's axis produces tilt error and the Z position error of a spot-face produces Z error.
\end{itemize}


Depending on the nature of these errors, we can distinguish
 those that can be compensated through software algorithms and 
 those that cannot be solved. 
 In the case of the DESI project the fiber positioning system benefits from
a fiber-view camera that can measure the position of each fiber 
(back-illuminated) after each movement.
 In this way, the main pointing algorithm executes sequentially an established number of iterations studying after each one the error between the real actual position and the theoretical target. Thus, the X-Y errors due to 1, 2 and 4 can be compensated with the corrections provided by the fiber-view camera system. However Z errors, tilt errors and the X-Y error due to 3 cannot be solved with this method.

\section{The positioner concept design}
\subsection{The mechanical design}

\begin{figure}[!h]
   \centering
   \includegraphics[width=7cm]{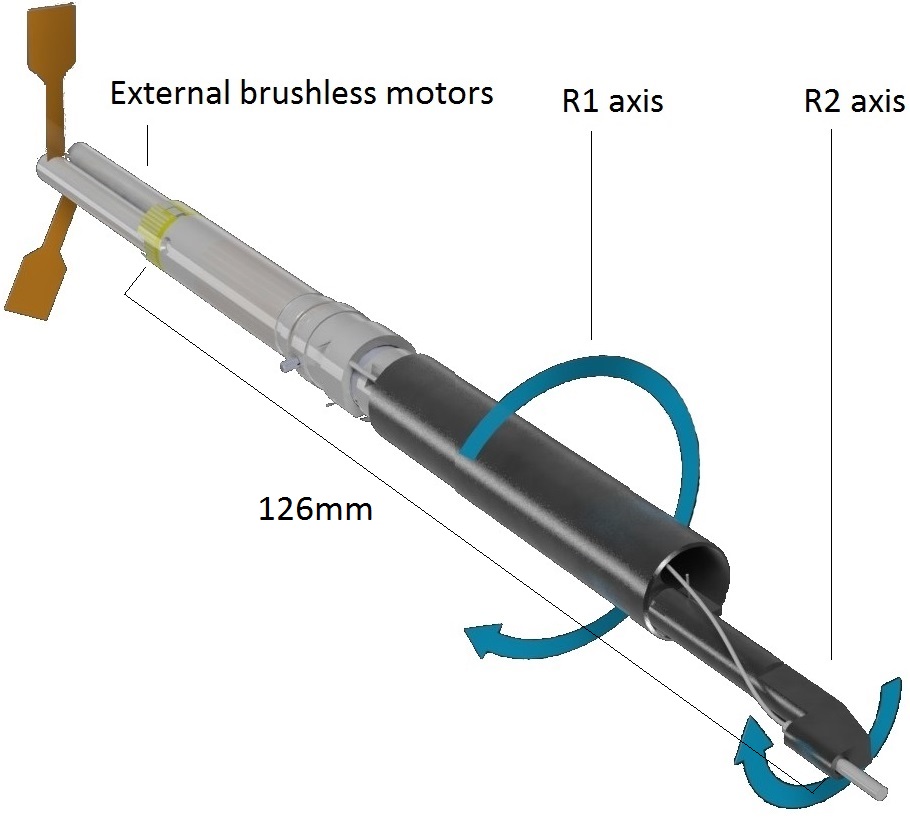}
         \caption{The new ES-CH 8-mm fiber robot postioner}
         \label{robot}
   \end{figure}

The basic mechanical design of the positioner was developed by the company AVS in collaboration
with the IAA-CSIC (see Fig.~\ref{robot}). Two prototypes were built by MPS taking into account a critical optimisation for serial massive production.  

\begin{figure}[!h]
   \centering
   \includegraphics[width=9cm]{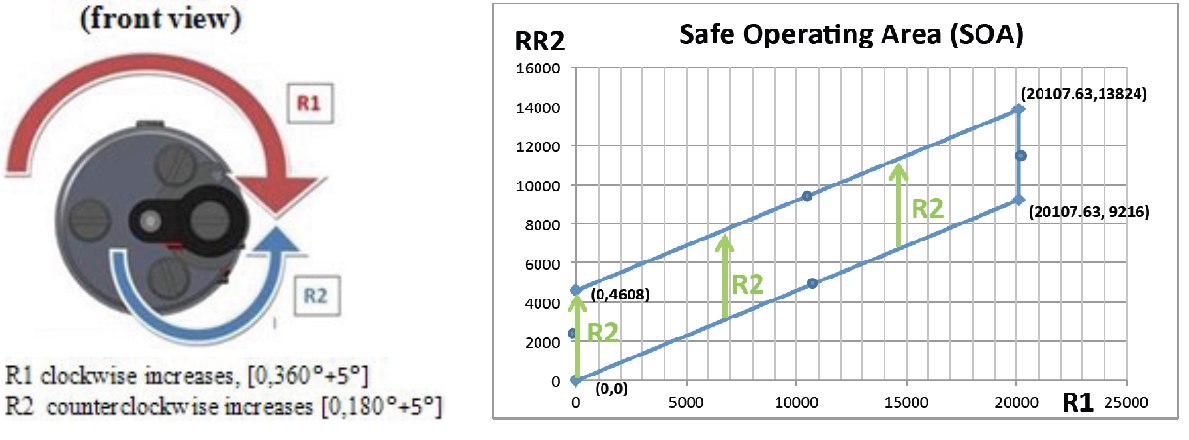}
      \caption{Left: schematic diagram of the positioner kinematic concept.
Right: Safe Operating Area (SOA) of the ES-CH positioner.
              }
         \label{ROT1ROT2}
   \end{figure}

The mechanical concept of the DESI 10.4mm-pitch fiber positioner presented in this paper
is based on two coupled rotations, one nested into the other. The rotation over the
main axis is called R1, which, in turn, houses R2. R2 is a rotation DoF nested off-axis
in the R1 mechanism. By the combination of R1 and R2 a grid of positions within the
Patrol Disc is provided. Figure \ref{ROT1ROT2} shows this kinematic concept.

A rotation in R1 induces a rotation of R2, which has to be compensated by the second motor. RR2 is the number of steps that the second motor has to do including this compensation. The Safe Operating Area (SOA) in steps is shown in Figure \ref{ROT1ROT2}, assuming 6 stable steps by turn of the motor (for use with digital hall sensors; using an encoder it is possible to increase the precision up to 110 steps per turn of the motor axis), and a gearbox of 1024:1 with R1=[$0-360^\circ$], R2=[$0-180^\circ$].

All the (R1,R2) to (R1, RR2) transformations are done inside the positioner control electronics. The positioner accepts Cartesian (X,Y) coordinates or their own native coordinates
(R1,R2), both coordinates system referred to its center. To do that, each positioner needs to know (L1,L2), where L1 and L2 are the lengths of the R1 and R2 arms. The positioner center and (L1, L2) are calculated with the help of the fiber-view camera.

The ES-CH fiber positioner includes the following main mechanical features:

\begin{itemize}
\item[1.]{\bf Preloaded transmission:} This improves dramatically the
repeatability performance of the positioner as the gearbox backlash is almost completely
cancelled.
\item[2.]{\bf Hard stops for both R1 and R2:} This prevents damage to the optical fiber in case
of incidental run-out of the motors.
\item[3.]{\bf Customised motor gearbox:} Decreased amount of parts to make the link with the main rotation axis, which implies an improved reliability of the device. This also leads to a lower cost of the fiber positioner.
\end{itemize}

The present concept based on fixed R1,R2 motors leads to the fact that a R1 motion induces a motion in R2, even without running the R2 motor. So, in order to avoid that, the induced gearing ratio has to be compensated by R2. If the fiber tip has to be moved by $\delta$R1 and $\delta$R2 to reach a final position, then R2 has to move $\delta$R2+($\delta$R1/2.181818...), where the latter figure corresponds to the induced ratio.

\begin{figure}[!h]
   \centering
   \includegraphics[width=7.5cm]{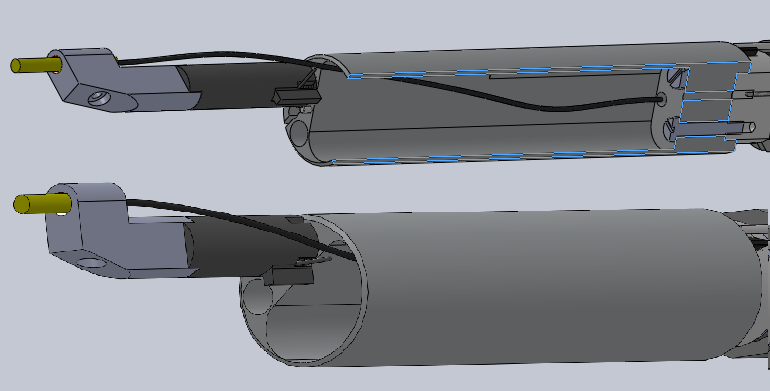}
      \caption{Fiber path in the ES-CH fiber positioner.
              }
         \label{fiber}
   \end{figure}

On the other hand, the optical fiber is attached to the ferrule at the arm tip, it makes an helicoidal spline up to a coaxial hole, and goes straight up to the exit by the rear side (see Figure \ref{fiber}). The minimal radius of curvature of the fiber along this path is 60-mm, so the present requirement given in Table 1 is fulfilled.

	Finally, after the evaluation of several driver technologies, a 4-mm DC brushless technology motor with hall-sensing control and integrated magnetic encoder, combined with a planetary gear-head, have been selected to ensure a reliable motion and precise accuracy. All necessary components were provided by FAULHABER-MINIMOTOR.

To fit the requirements of the positioner in terms of dimensions, a "watch design and
manufacturing technology" have been considered to miniaturise the mechanical parts such
stator, rotor, bearing, gears, shaft, etc. To ensure the accuracy of the systems and quality control, all components were manufactured with automatic CNC machines in house at MPS. Table \ref{motors} shows the main characteristics of the motors.

\begin{table}[!htt]
\begin{tabular}{l l l}
\hline
{\bf Dimensions}& \\
\hline
            Diameter: & 4 mm \\
	Overall length: & 34 mm  \\
	\hline
{\bf Performance}	& \\
	\hline
	Motor speed: & 30,000 rpm @ 3.3 V\\
	Output torque of the motor: & 0.0108 mNm\\
	Sensor:	& from 6 to 128 pulse per rotation.\\
	Gear Ratio: & 1024:1 \\
	Output torque of the gear-motor: & 10 mNm\\
     \hline
						
         \end{tabular}
\caption{Main  characteristics of the Faulhaber 4mm gear-motor.}
         \label{motors}
   \end{table}
\subsection{Control electronics}

In order to move the two motors involved in the positioner, and optimising its performance, a basic electronics has been developed. This electronics will be placed inside the chassis of the robot. On the one hand, the components that conform the control
electronics have been chosen following ultra-low power consumption requirements; this
characteristic allows to minimize the thermal noise in the focal plane. On the other hand, all electronic components have to respect the volume constraints required by the board dimensions.

\begin{figure}[!h]
   \centering
   \includegraphics[width=8.5cm]{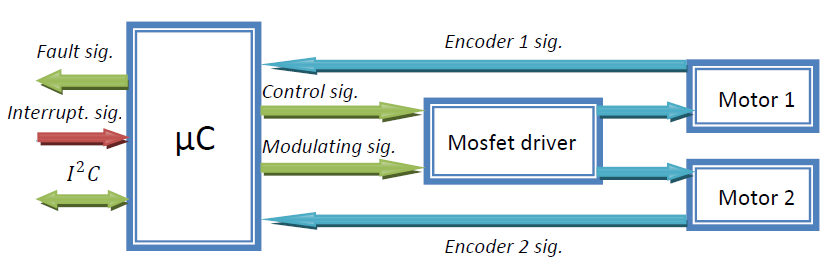}
      \caption{Overall control electronics block diagram.
              }
         \label{electronics}
   \end{figure}

Figure \ref{electronics} shows the overall control electronics block diagram. The main component of the system is a 32-bit micro-controller, in charge of receiving commands from the top communication nodes, processing them and managing information flowing into the positioner. This component will receive data through a fast $I^2C$ bus up to 100MHz, and will control with high accuracy the position of each motor through a full mosfet-based bridge system. The real position of each motor will be provided to the micro-controller thanks to a magnetic encoder system embedded in each motor. This information will allow the microprocessor to modulate in real time the velocity of each motor, accomplishing the time constraints demanded by the Master system in order to follow the collision avoidance algorithm. On the other hand, this information will allow the system to know the existence of real movements and registering each time the existence of movement errors (hard motor fault, arrival to a physical zero, etc).

To optimise the consumption of the electronics, the micro-controller can run in two modes, i.e. Active and Sleep. The first one is used to process data, to monitor constants and to
position the robot, whereas the second one is used in waiting command periods, reducing
consumption to a minimum when not required. This feature is managed by an interrupt
line that allows the Master software to manipulate directly the consumption state of each
positioner.

Finally, the motor control algorithms involved in our control electronics allow us to manipulate in real time the current that will flow though each motor, configuring directly the torque that the positioner will use to move the fiber. This feature allows to prevent some no movement motor errors caused by mechanics stress. To accomplish this goal, the
magnetic encoder system will be used to check the existence of real movements and a
windowed torque in case of the existence of this kind of errors.

 Tables \ref{tabla_electronica} and \ref{tabla_low} show the main performance characteristics of the developed control electronic board. 
 
\begin{table}[!htt]
\begin{tabular}{l l l}
\hline
{\bf Characteristic}&{\bf Value}\\
\hline\hline
Power supply& 3.3 v\\
\hline
Electrical power& Motor on: 1204.5mW\\

& Receiving command:79.2mW\\

& Sleeping: 0.66$\mu$W\\
\hline
Connector Pin  & 1 +3.3v\\

& 2 ground\\

& 3 SDA\\

& 4 SCL\\

& 5 Fail/OTB program\\

\hline
Mobile cables inside the positioner& No\\
\hline
Temperature housekeeping& Yes\\
\hline
Stop error sensed& Hall sensors detect \\
& bad stop condition \\
\hline
\end{tabular}
\caption{ES-CH positioner electronic characteristics}\label{tabla_electronica}
\end{table}


\begin{table}[!htt]
\begin{tabular}{l l l}
\hline
{\bf Characteristics}&{\bf Value}\\
\hline\hline
Sleep motor& Yes\\
\hline
Sleep microcontroller&Yes\\
\hline
X-Y to R1-R2 bidirectional transformations&Yes\\
\hline
OTB programmable&Yes\\
\hline
Error management & If motor is blocked \\
&then stop positioner\\
& and send an error \\

\hline
\end{tabular}
\caption{ES-CH positioner low-level software characteristics}\label{tabla_low}
\end{table}

It is important to highlight that the reconfiguration time of the positioners is a main variable that allows to maximise the number of spectra taken in an observation night. This parameter is defined as the time that the positioner takes to move the fiber head and to compensate the placing errors after the execution of the fiber view camera system re-configuration algorithm (see section \ref{testing}). The control electronics developed for our prototype allows to move the robot from end-to-end of the parol disk in up to 35s, whereas the time taken to execute up to five iterations of the correction algorithm is 60s. This last period includes the time required to get the images, to process them in order to measure the error, and to move the robot to the correct place. However, this reconfiguration time does not have a major importance in this moment for this prototyping phase, but it will be optimized as a characteristic constant in the final system.

\subsection{Communication solution}

We adopted a wired solution for the communications of the DESI positioners. Following this scenario; all placement orders, configuration messages and alert advices will be sent from superior communication nodes to each positioner through a fast plus I2C bus achieving transfer rates up to 3.4Mbps. This serial communication solution will route all the data packets from the master computer to all the positioners through two router level modules. These modules will allow the distribution of the targets on sectors, respecting the maximum
capacitance charges. This protocol allows to link per unit and distributing the overall system workload.

\begin{figure}[!h]
   \centering
   \includegraphics[width=8.5cm]{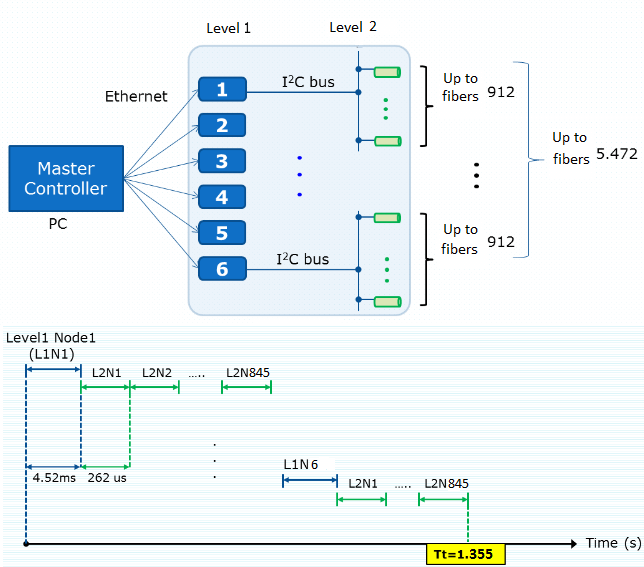}
      \caption{Up: Overall communication architecture based on fast $I^2C$ plus Ethernet. Down: timing simulation example.
              }
         \label{comm1}
   \end{figure}

On the other hand, in order to avoid the relevant bottle neck in the first router level,
this link is replaced by an Ethernet connection that allows data transferences up to 1000Mbps. 
Making a standard interface between the master computer and the first I2C components, this structure allows different load distributions in order to respect the focal plate structure.
Figure \ref{comm1} shows the overall communication proposal for DESI. This proposal allows the master computer to link with all the positioners distributing the workload among the three router levels. On the other hand, it shows the time taken by the communication system to route one position message (3 bytes) to the whole system.

Finally, continuing with a low power consumption philosophy in order to minimise the thermal
noise in the focal plane, this architecture has been designed selecting ultra-low power
consumption devices. Achieving power consumption levels up to 500mW at 3.3V for all
branches pending of each level 1 nodes.

\subsection{Collision avoidance algorithm}\label{algorithms}

To get the most efficient scientific results from the planned survey, the 5,000 fiber positioners need to move the fiber ends in parallel. As they share workspace, the most direct
trajectories are prone to collision that could damage the robots and certainly have a serious
impact on the survey operation.

\begin{figure}[!h]
   \centering
   \includegraphics[width=9cm]{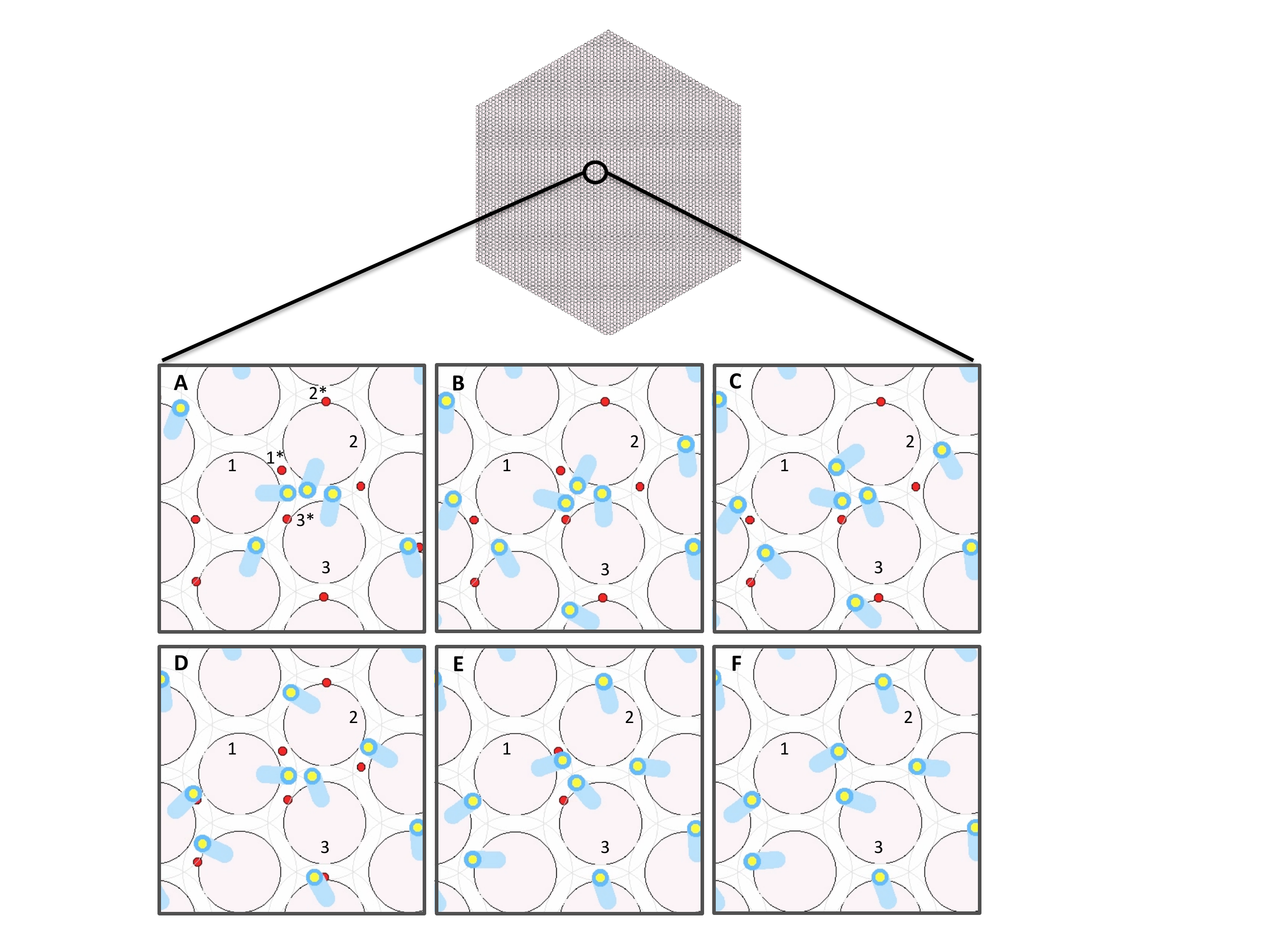}
      \caption{Simulation environment for 5,000 positioners and a zoom on the motion of three
positioners. The six boxes (A to F) show six snapshots of the simulation. 1*, 2* and
3* are respectively the target positions for positioner 1,2 and 3. These three positioners
are engaged in a local conflict in which positioner 1 needs to make space for positioner 2
to pass. Positioner 2 cannot make room for positioner 1 because positioner 3 is blocking
the way. Positioner 3 needs to pass both positioners to reach its target point. The small operation from positioner 1 moves this positioner farther from its target point but it makes
room for the positioner 2 to pass safely. When positioner 2 clears the way, positioner 1
starts moving toward its target point and this gives a safe way to the positioner 3.
              }
         \label{collision}
   \end{figure}

As the first step, we have developed a visual simulation environment for the whole 5,000 positioners for testing and validation of target assignment (see \cite{Morales2012}), motion planning and collision avoidance algorithms (See Figure \ref{collision}).

   Then, we have provided a motion-planning algorithm for motors. The goal is to move in
parallel and in minimum time all positioners that are initially packed at a focal plane.
The main challenges are ensuring collision avoidance and minimising reconfiguration time.

Our proposed motion-planning framework for DESI is based on a novel decentralised navigation
function. The navigation function takes into account the configuration of positioners as
well as the positioner constraints. An important aspect of our design is that we provided
proof of convergence and collision avoidance. Decentralisation results in linear complexity
for the motion planning as well as Independence of motion duration with respect to
the number of positioners. Therefore the coordination method is scalable for large-scale
spectrograph robots like DESI. The short in-motion duration of positioner robots, approximately
2.5 seconds in DESI, will thus allow the time dedicated for observation to
be maximised. Upon the preparation of the control board and positioner prototypes, the
designed trajectories will be tested in a close loop with the motors. The velocity profiles
corresponding to the configuration presented in Figure \ref{collision} and more detailed information can be found in (\cite{Makarem2014}).

\section{$\O$4mm DC-brushless motors performance}
In order to ensure that the motors meet the expected specifications in terms of accuracy, repeatability, and power consumption, three different tests have been developed. 

\subsection{Backlash and repeatability}
\begin{figure}[h]
   \centering
   \includegraphics[width=8.5cm]{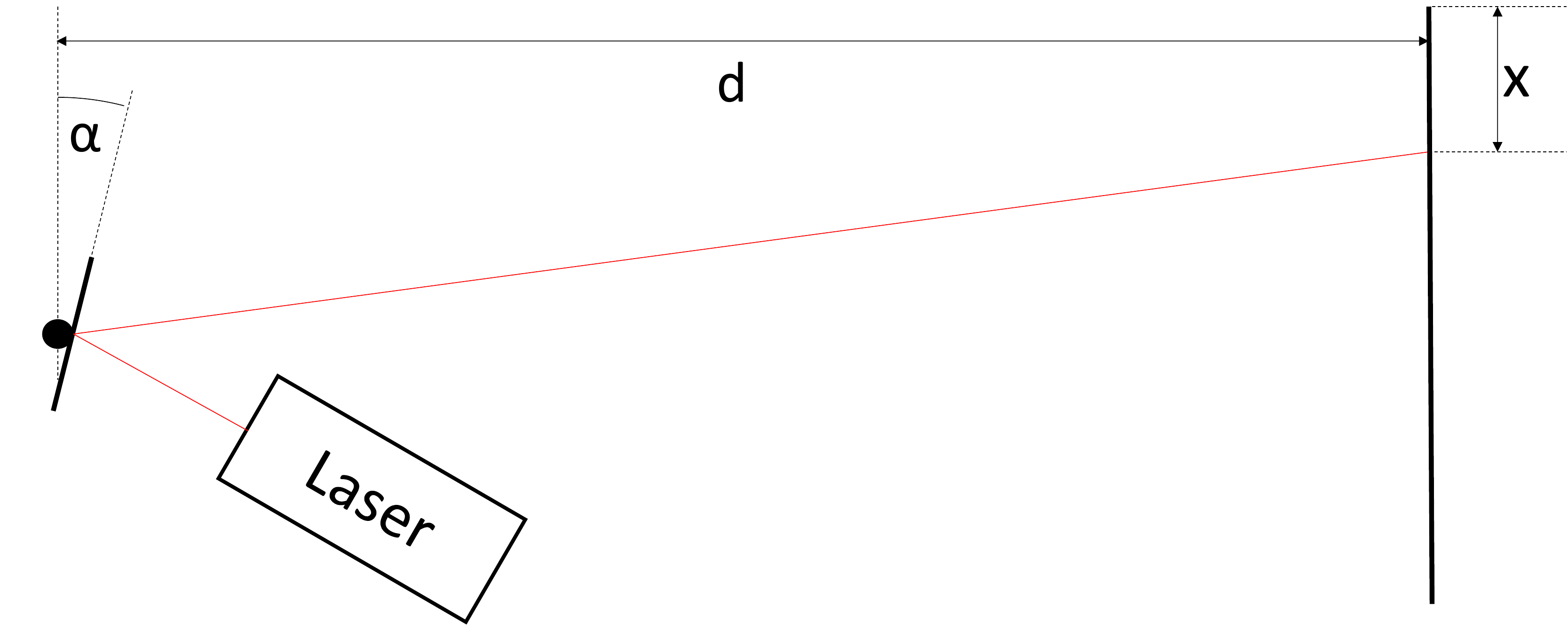}
      \caption{Schematic view of the measurment set up.}
         \label{mirror}
   \end{figure}

To study the performance of the DC motors, a small mirror is mounted on the output shaft and a laser is pointed on the mirror (see Figure \ref{mirror}). The reflected laser beam illuminates a spot on a wall. The position of the spot is related to the angle of the output shaft by $x = 2 \alpha d$, where x is the position of the laser spot on the wall in millimetres, $\alpha$ is the angular position of the output shaft in radians and $d$ is the distance between the mirror and the wall. In our case $d = 7000 mm$.

The backlash and repeatability are measured according to the following procedure:
   \begin{itemize}
   \item[1.]Move +X steps, then do -X steps
   \item[2.]Measure position of the output shaft at position 0
   \item[3.]Move -X steps, then do +X steps
   \item[4.]Measure position of the output shaft at position 0
   \item[5.]Go to 1.
   \end{itemize}
   The measuring unit is $60\deg$ which defines one step on the motor shaft and corresponds to one increment of the Hall sensor.

The Measurements at point 2 represent the position of the output shaft
when coming from one side while the measurements at point 4 represent its position
when coming from the other side. In both measurements the position of the motor is
the same, but the backlash interferes differently from the two sides. The difference between the
measurements at points 2 and 4 represents the backlash in the reduction gears. The deviation
within the measurements at point 2 represents the repeatability of positioning. The same applies for the measurements at point 4. 

\begin{table}[h]
\centering
\begin{tabular}{l l l l l}
\hline
{\bf Device} & {\bf Repeatability -} & {\bf Repeatability +} & {\bf Backlash} \\
 \hline\hline
Motor 1 & 0,076 & 0,288 & 0,043 \\\hline
Motor 2 & 0,229 & 0,161 & -0,006 \\\hline
Motor 3 & 0,136 & 0,034 & 0,741 \\\hline
Motor 4 & 0,212 & 0,212 & 0,928 \\\hline
Motor 5 & 0,042 & 0,042 & 3,151 \\\hline
Motor 6 & 0,424 & 0,068 & 0,430 \\
\hline
\end{tabular}
\caption{Numerical results [deg] of tests with 500 steps at 1000 rpm}
\label{motor1}
\end{table}

Table \ref{motor1} summarises the results of our measurements on six custome-made motors moving $500$ step  at the speed of $1000$~$rpm$. The first observation is that Motor 5 has a relatively large backlash. This is probably due to an imperfection or dust in the reduction gears. The backlash of the other motors stays in an acceptable range. All the motors show satisfactory repeatability according to the technical requirements.


It is important to note at this point that these measurements include only the motor and gearhead and that the DESI ES-CH fiber positioner includes preloaded springs to cancel backlash in both rotations.

\subsection{Power consumption}

In these series of measurements, the motors are driven at a constant speed in both directions in a closed loop using the digital hall sensors. An amperemeter measures the electric current passing through the power bridge that drives the three phases of the motor.

\begin{figure}[h]
\begin{center}
\includegraphics[width=9cm]{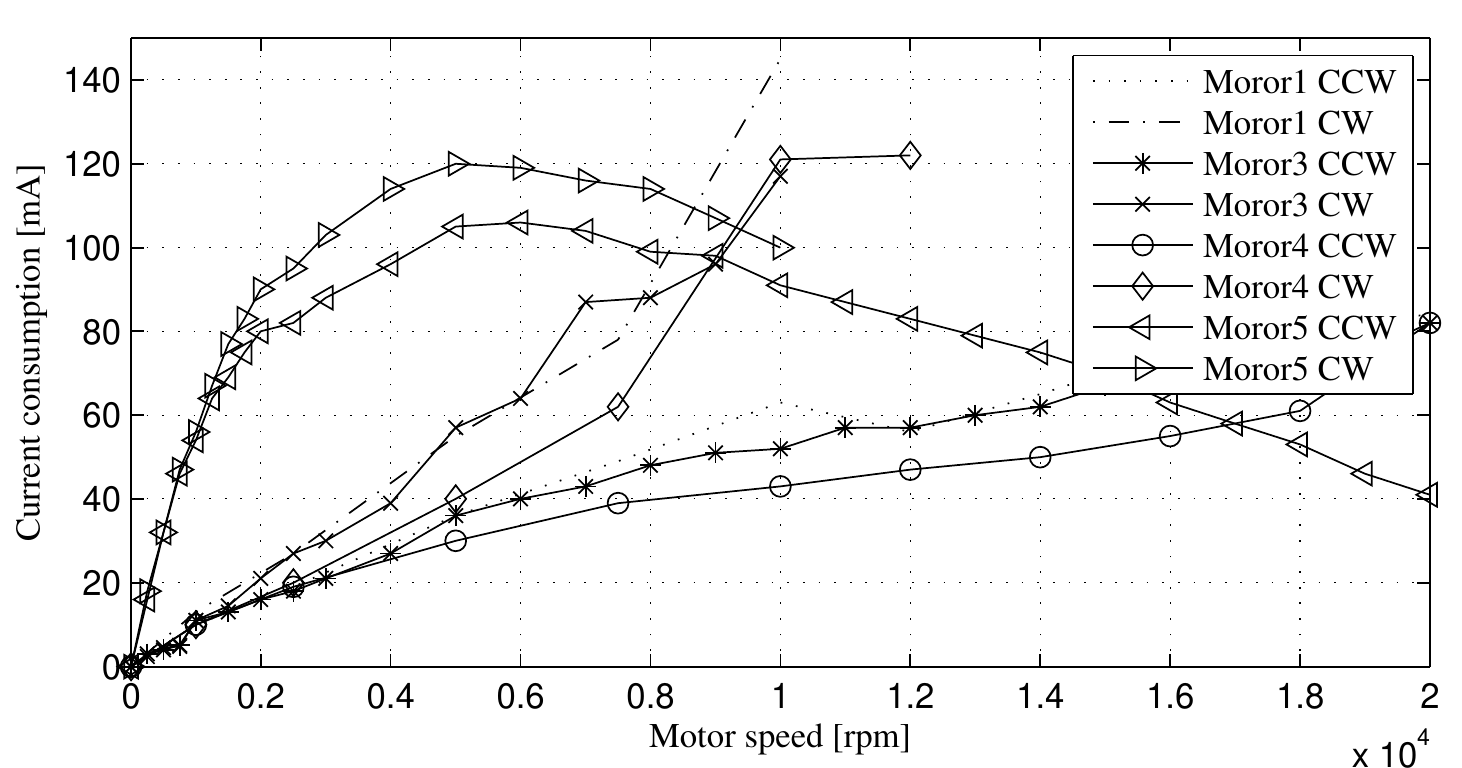}
\caption{No load current consumption of Motors 1,3,4, and 5 in both directions.}
\label{CurrentConsumption}
\end{center}
\end{figure}

Figure \ref{CurrentConsumption} shows the current consumption of Motors 1, 3, 4, and 5 in clockwise and counter-clockwise rotations.

The first observation is that motor 5 has an unexpected non-linear behaviour. Its current consumption has a maximum at about 5000 rpm. Then it decreases toward the maximum speed. This can be explained by the manufacturing defect which could be also responsible for the big backlash and seems to have less influence at higher speeds. 

The second observation is that the other motors rotating clockwise have a consumption about twice as high as rotating counter clockwise (CCW). In addition, the maximum speed of clockwise rotation is about half of the counter clockwise maximum speed. This is due to the fact that the angular offset between the motor and the hall sensor is optimized for counter clockwise rotation. In the future we will set the offset as a function of turning direction and expect the same consumption in both directions. In case of using encoders instead of hall sensors to provide feedback, the consumption can be reduced by another $5\%$. 

The current consumption of two motors plus the electronics will easily meet the requirements of maximum heat dissipation. The current measured here times the operating voltage (3.3V) gives the power consumed by one motor including dissipation in the power bridge. With the requirement of max 1.2W of heat dissipation per positioner, the requirement is 363mA per positioner. In the worst case (both motors at full speed) our positioner consumes only 160mA.

\subsection{Current consumption control}

Knowing the minimum torque that moves the motors is needed not only for minimizing the consumption of the whole system both also to have total control of the positioner movement. Therefor we conducted a test for measuring the minimum torque for the two of the motors (Motor 1 and Motor 2).

It is worth mentioning that the electronics developed specifically for the fiber positioner are able to control the maximum current that will pass through each motor (therefore the torque given to each one), by modulating the control lines with a PWM signal. Thus, the torque of each motor is proportional to the duty cycle of the modulating PWM signal.

In this test, the duty cycle of the modulating signal has been incremented from 1\% up to 50\%, measuring the current consumption while the movement of the motor is measured by the Hall sensor.

\begin{figure}[htb]
\centering
\includegraphics[width=9cm]{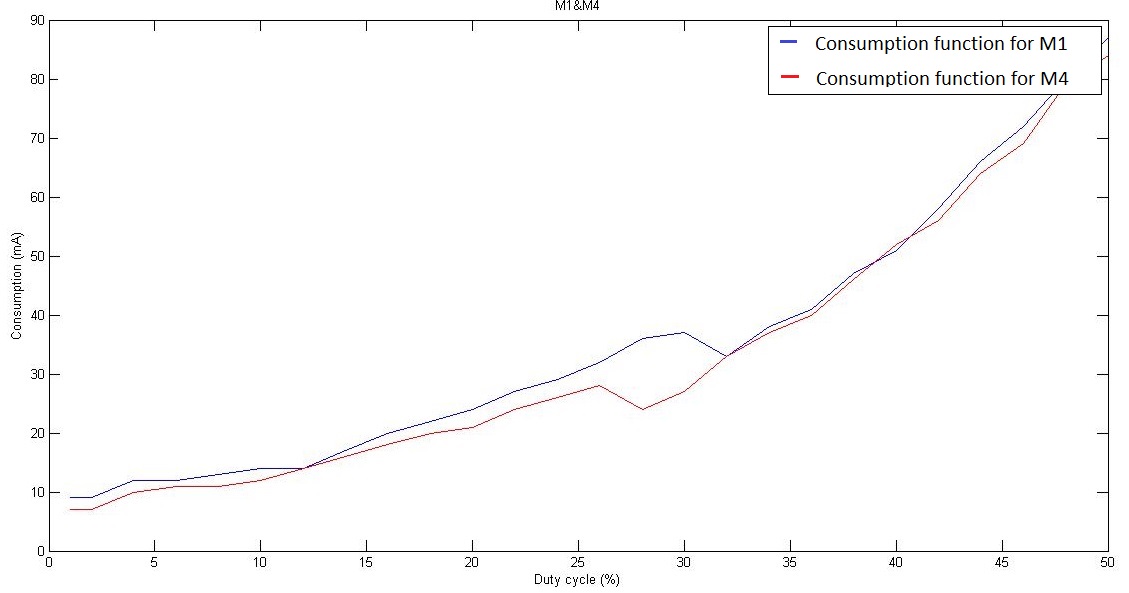}
\caption{Motor 1 and Motor 2 consumption comparative.} \label{motor3}
\end{figure}

Table \ref{motor2} and Figure \ref{motor3} show the results achieved in this test. The Figure shows the consumption of the motors as function of the duty cycle. The duty cycle from which the motor starts a bad movement (vibrations) corresponds to the point where the consumption begins to decrease. The duty cycle from which the motor starts to move correctly is the point where the consumption starts to rise again. 
Looking at the consumption where the motor moves correctly, we can observe that it is linear and very similar for both motors. Thus we can conclude that the minimum torque that we can applied to each motor in order to control them correctly is 32\% of the maximum.

\begin{table}[h]
\centering
\begin{tabular}{c c c c}
\hline
{\bf Motor} & {\bf Starting (non-regular)} & {\bf Fails} & {\bf Starting (regular)}\\
\hline \hline
Motor 1 & 16 & 24-30 & 32   \\
\hline
Motor 2 & 14 & - & 27  \\ 
\hline
\end{tabular}
\caption{Movement results showed through duty cycle ranges.}\label{motor2}
\end{table}

\subsubsection{Magnetic coupling}

The two motors involved in each robot are installed in parallel positions with a considerably short distance. Due to the magnetic principles that control this kind of motors, it is important to study the effect of the magnetic field of one motor on the movement of the other (inductive effects). In addition, the magnetic field could cause noise and distortions on the Hall sensor signals.

In order to study the relevance of magnetic effects in the motor performance, we have measured the emitted magnetic field of one motor in movement with an analogical Hall sensor, the electric induction in one disconnected motor winding and we studied the coupling between two running motors.

To prove the real magnetic coupling influence in the motor control, the two motors are moved the same number of steps, one after the other and the total number rising edges of the magnetic encoder signal of each motor are compared.
After a significant set of tests  it has been proven that there is no influence in the individual movement of the motors when they are in direct contact.

\section{Positioner testing set-up}

\label{testing}

A camera-based fiber view set-up has been used to characterise the performance of our positioner. This set-up allow us to ensure that the DESI requirements listed in Table 1 are met in terms of accuracy, backlash and repeatability. The multi-iteration procedure for high precision positioning has also been evaluated.

\begin{figure}[!h]
   \centering
   \includegraphics[width=8cm]{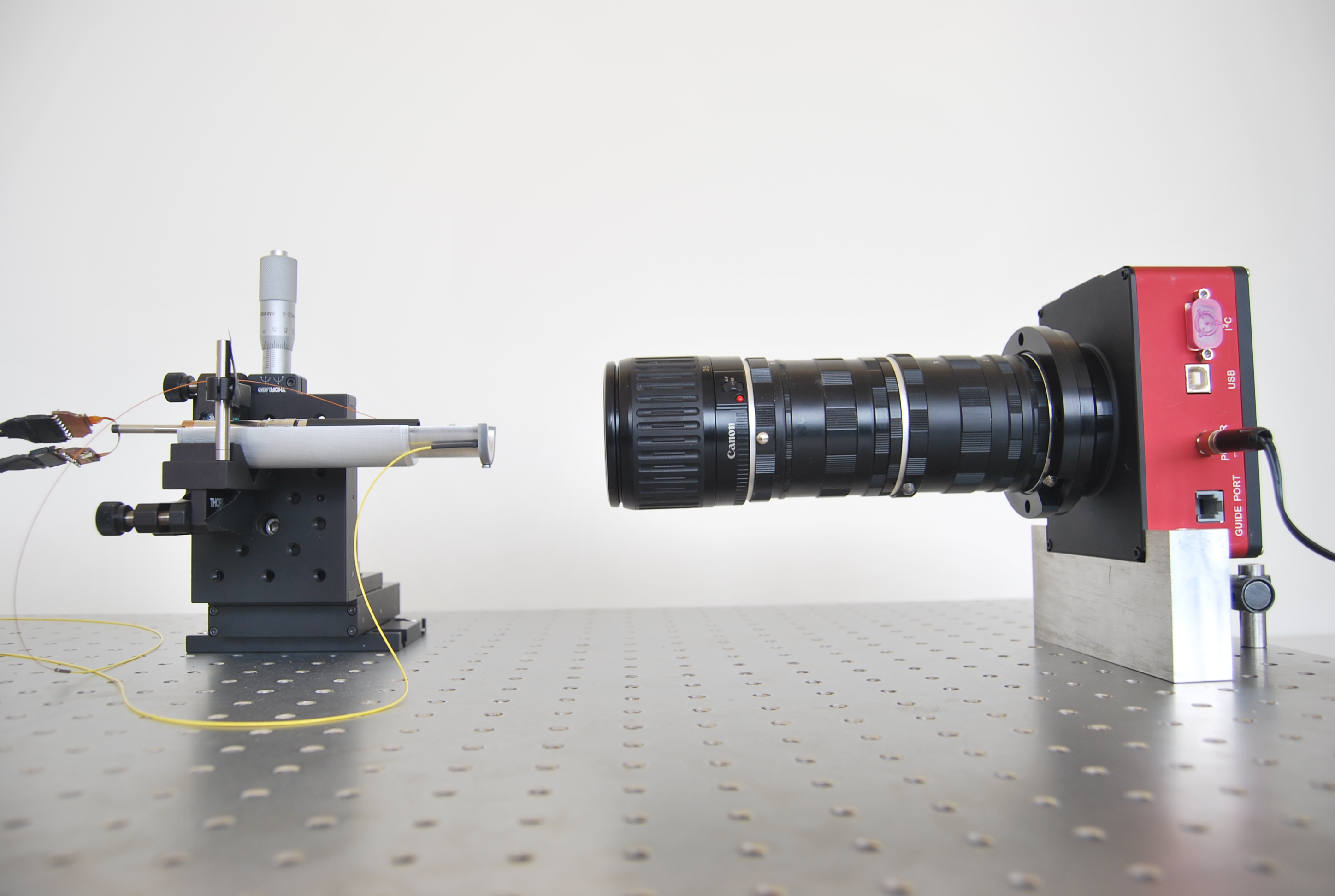}
      \caption{Lateral view of the optical set-up.}
         \label{camera}
   \end{figure}
 
 The optical set-up setup consists on a multi axis positioning stage holding the fiber positioner, and a Santa Barbara CCD camera developed for astronomical applications. This camera is installed in front of the actuator. To minimize the vibrations and guarantee the stability, all these components are mounted on an optical granite table and assembled using specific aluminium parts.

Figure \ref{camera} shows a lateral view of the optical setup with
the positioner placed in front of the camera. As hown in the figure, the CCD camera has been mounted with two macro expanders and a Cannon macro lens.

The camera is connected to a PC where an automatic Matlab routine developed by J. Silber at LBNL is executed. This routine communicates with the user interface of the positioner, it sends a position target, wait for the end of the movement, and takes a picture. After each shot, the routine extracts the distance between the fiber moved by the positioner and a second  optical fiber used as a reference. 


It then computes the difference between the fiber position 
and the theoretical one, establishing this as the system accuracy. On the other hand, this system allows to calibrate the positioner at the beginning of each test, establishing the actuator center, the real length of its arms (L1, L2), and configuring the allowed positions.

\begin{figure}[!h]
   \centering
   \includegraphics[width=8.5cm]{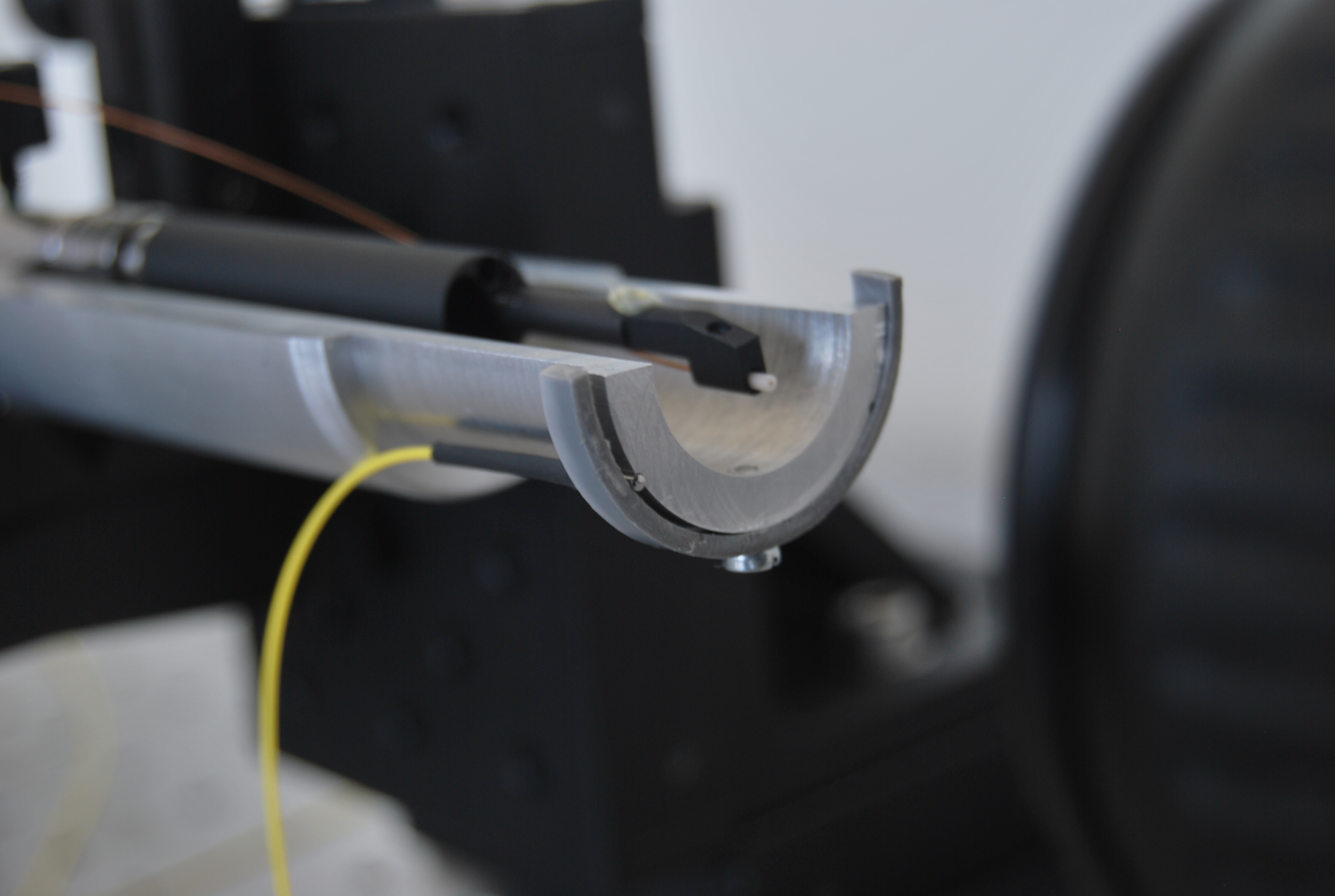}
      \caption{Reference fiber holder.}
         \label{reference}
   \end{figure}
   
Figure \ref{reference} shows the reference static fiber mounted in an aluminium piece built for this purpose.

The Matlab program executed in this optical setup emulates the main actions that the master computer of the DESI telescope will have to perform once the 5000 positioners will be mounted on the focal plate.
To maximise the accuracy, this routine executes a pre-established number of iterations, determining after each movement the error between the positioned fiber head and the static reference, using this data as the input of the next iteration in order to make the necessary corrections. The X-Y accuracy results obtained will be shown as the main representative performance data for the characterisation of the ES-CH 8-mm positioner.

\section{Results}

The three main variables that characterise this kind of $\theta - \phi$ positioner, are the repeatability, the accuracy and the mechanical hysteresis. These data ensure that the positioner will place the fiber head at the same point in extremely different situations, minimising the error between the theoretical target and the attained one.  

\subsection{Accuracy measurements}

In order to study the accuracy of the positioner, a simple test has been developed. A set of position targets evenly distributed on a grid and filling the whole patrol area is defined. The master routine commands the positioner to move successively on every target point executing five correction iterations for each evaluated position.

\begin{figure}[!h]
   \centering
   \includegraphics[width=8.5cm]{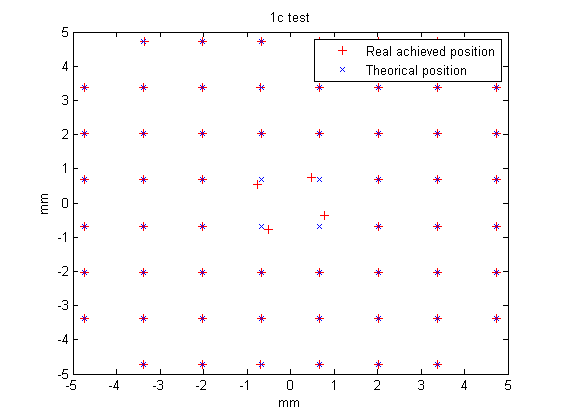}
      \caption{Results of one grid test executed with the optical set-up.}
         \label{test1}
   \end{figure}

Figure \ref{test1} shows the results obtained with one of the positioner prototypes. This figure presents the theoretical positions (blue crosses) and the achieved ones (red crosses).

\begin{figure}[!h]
   \centering
   \includegraphics[width=8.5cm]{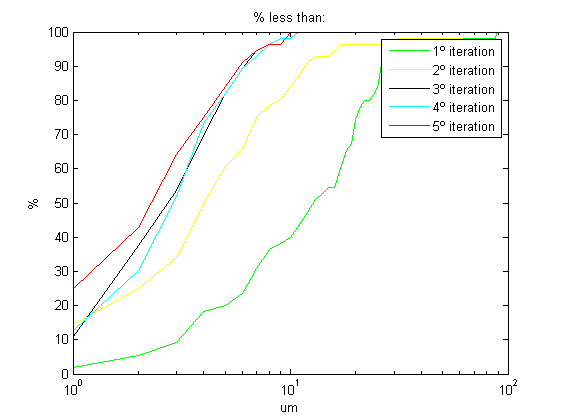}
      \caption{Iterations absolute x-y error evolution.}
         \label{test2}
   \end{figure}

The main purpose of this procedure consists on obtaining the maximum number of target positions with an absolute x-y error less than 5$\mu$m (value obtained with the previous 10-mm diameter prototype developed by the Spanish group for the former BigBOSS project (\cite{Fahim2013})). 
Figure \ref{test2} shows the effect of the
corrective iterations on the position accuracy. This preliminary result
shows that 80\% of the points can be reach with an absolute error less than 5$\mu$m
after five iterations of the corrective algorithm.

These results have a significant relevance, because all of them have been obtained without running any hysteresis reduction algorithm, see below. In fact, all the results collected in figures \ref{test1} and \ref{test2} meet the requirmenents of the DESI project in terms of maximum absolute error, RMS error and percentage of positions with less than 5$\mu$m in absolute x-y error. 

Table  \ref{tab_results2} shows the data collected in these preliminary tests. It is possible to observe that the maximum absolute error after the fifth iteration is lower than 10$\mu$m (maximum value allowed, see Table 1). On the other hand it is important
to highlight that the resulting RMS error is less than
5$\mu$m after the fifth iteration.


\begin{table}[h]
\centering
\begin{tabular}{c c c c c c c}
\hline
&{\bf $<$5$\mu$m } &{\bf $<$2$\mu$m  }&{\bf $<$1$\mu$m  }&{\bf  RMS  }&{\bf  Max  }&{\bf Min} \\
\hline\hline
No Iter.&   1.8\%    & 0\%   &  0\%   &  138.1$\mu$m& 342.2$\mu$m & 3.4$\mu$m\\
1º Iter.&  20.0\%    & 5.5\% &  1.8\% &  19.9$\mu$m& 88.1$\mu$m & 0.5$\mu$m\\
2º Iter.&  60.7\%    & 25.0\%&  14.3\%&  11.1$\mu$m& 62.6$\mu$m & 0.4$\mu$m\\
3º Iter.&  82.1\%    & 37.5\%&  10.7\%&  3.9$\mu$m&  10.6$\mu$m& 0.3$\mu$m\\
4º Iter.&  82.1\%    & 30.4\%&  12.5\%&  3.8$\mu$m& 10.4$\mu$m & 0.5$\mu$m\\
5º Iter.&  83.9\%    & 42.9\%&  25.0\%&  3.6$\mu$m&  9.8$\mu$m& 0.2$\mu$m\\
\hline

\end{tabular}
\caption{Main absolute x-y error results.}\label{tab_results2}
\end{table}


\subsection{Backlash characterisation }

In this subsection, we aim to quantify the cumulative effect of the different transmission stages on the final positioning accuracy. 

From the results obtained with the optical set-up it was possible to characterise the internal mechanical response of the evaluated prototype. In particular, one of the most representative mechanical characteristic of this robot is the internal hysteresis. The position of the fiber depends on the movement of two motors.
The backlash of each rotation axis depends on the direction of rotation of the motor, the amplitude of the displacement and the velocity profile. Thus, the positions achieved by the robot while trying to reach one target in the patrol disc will be different for every starting point.

\begin{figure}[!h]
   \centering
   \includegraphics[width=8.2cm]{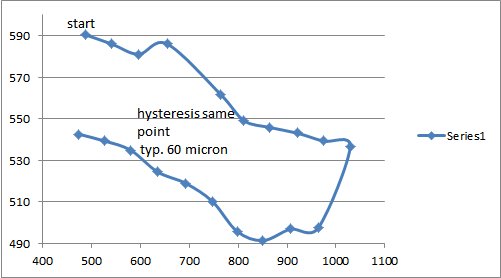}
      \caption{Characteristic hysteresis pattern in the x-y plane (units are $\mu$m).}
         \label{hysteresis}
   \end{figure}

Figure \ref{hysteresis} shows the results obtained while trying to move from one starting point (A) to a target point (B), and returning to the starting point. A characteristic mismatch can be seen between the points along the path from A to B and the points along the path from B to A.

In order to minimise this mechanical error, a software algorithm able to reduce up to ten times this error was developed. To make it real, this algorithm forces the robot to move inside a preliminary backlash region, before it reaches the target position. This action softens the internal mechanics before the final movement. It is important to note that the smaller the robot, the more difficult it will be to reduce this kind of mechanical errors due to the soft properties of the metals of these small sizes.

\begin{figure}[!h]
   \centering
   \includegraphics[width=8.2cm]{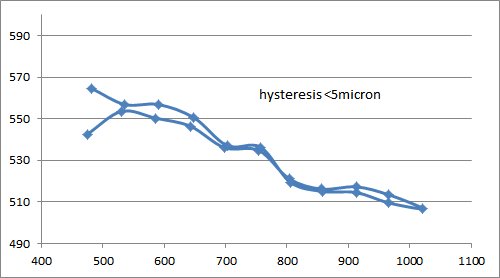}
      \caption{Hysteresis pattern after applying the hysteresis reduction algorithm.}
         \label{hysteresis2}
   \end{figure}

Figure \ref{hysteresis2} shows the results of the test run shown in figure \ref{hysteresis}  after the implementation of the hysteresis reduction algorithm.

The hysteresis errors depend only on the internal mechanic and physical behaviour, thus the algorithm developed to reduce them has a specific design for each fiber actuator and its particular
characteristics.

Finally, table \ref{tab_results_summarize} summarises the requirements for the DESI fibre positioner and the results obtained by the ES-CH fiber positioner. 

\begin{table}[h]
\centering
\begin{tabular}{l l l}
\hline
            {\bf Item}                  & {\bf Required}            & {\bf Achieved} \\
\hline\hline
            RMS XY Error                & $5\mu$m                 & $3.6\mu$m \\
            Max XY Error                & $10\mu$m                & $9.8\mu$m \\
            Z max defocus error         & $30\mu$m                & $6.4\mu$m \\
            Max angular tilt            & $\pm 0.1{}^{\circ}$       & $\pm 0.058^{\circ}$ \\
\hline
            Max reconfiguration time    & $20$sec                   & $30$s \\
			Power while active          & $< 1.204$W                & $0.6$W \\
			Power while sleep           & $< 0.2$mW                 & $0.66 \mu$W \\
            Fiber bend radius           & $> 50$mm                  & $60$mm \\
            Mass of the positioner      & $< 50$g                   & $30$g \\
\hline
\end{tabular}
\caption{Requirements and achieved values.}\label{tab_results_summarize}
\end{table}

\section{Conclusions}

We have presented here the  development of a new
Spanish-Swiss (ES-CH) 8-mm fiber robot positioner that 
fulfils all the requirements of the DESI project.
The main features of the mechanical design of our positioner are:
i) the two motors are located at the bottom
of the positionner and no electric wire
circulate through the positioner;
ii) the fiber path is relatively straight, ensuring
minimal light loss.

We have also designed a control scheme to minimise the electronics consumption, maximising the system accuracy and defining new command algorithms in order to avoid any hypothetical collisions between adjacent positioners. 
On the other hand, failure alert techniques developed in the control electronics ensure the avoidance of mechanical critical failures, lengthening the whole system life and minimising the need for maintenance. 

Based on the different tests performed,
we believe that our development has lead to
one of the most robust fiber robot positioner for massive spectroscopy instruments. The technology developed
may possibly be adapted to other multi-fiber
spectrograph projects.

\begin{acknowledgements}
We acknowledge support from the Spanish MICINNs Consolider-Ingenio 2010 Programme under grant MultiDark CSD2009-00064, HEPHACOS S2009/ESP-1473, and MINECO Centro de Excelencia Severo Ochoa Programme under grant SEV-2012-0249. We also thank the support from a CSIC-AVS contract trough MICINN grant AYA2010-21231-C02-01, and CDTI grant IDC-20101033; and support from the Spanish MINECO research grants AYA2012-31101 and  FPA2012-34694. JPK, PH and LM acknowledge support from the ERC advanced grant LIDA and from a SNF Interdisciplinary grant.
We thank J. Silber and R. Bessuner at LBNL for helping to set-up a replica of their testing optical bench set-up at EPS-UAM.
\end{acknowledgements}

\begingroup
\bibliographystyle{jtbnew}
\bibliography{bibliografia}
\endgroup

\end{document}